\documentclass[aps,prab,twocolumn,superscriptaddress,longbibliography]{revtex4-2}

\usepackage{units}
\usepackage{subcaption}
\usepackage{amssymb}
\usepackage{amsmath}
\usepackage{graphicx}
\usepackage{float}
\usepackage{dcolumn}
\usepackage{bm}

\begin{document}

\preprint{APS/123-QED}
\title{Seeding of Self-Modulation using Truncated Seed Bunches as a Path to High Gradient Acceleration}
 
%

\author{N.~Z.~van~Gils}
\affiliation{PARTREC, UMCG, University of Groningen, Groningen, NL}
\affiliation{CERN, 1211 Geneva 23, Switzerland}

\author{E.~Belli}
\affiliation{CERN, 1211 Geneva 23, Switzerland}
\affiliation{John Adams Institute for Accelerator Science, University of Oxford, Oxford, UK}

\author{M.~Bergamaschi}
\affiliation{CERN, 1211 Geneva 23, Switzerland}
\affiliation{Max Planck Institute for Physics, Munich, Germany}

\author{A.~Clairembaud}
\affiliation{Max Planck Institute for Physics, Munich, Germany}

\author{A.~Gerbershagen}
\affiliation{PARTREC, UMCG, University of Groningen, Groningen, NL}
\affiliation{John Adams Institute for Accelerator Science, University of Oxford, Oxford, UK}

\author{E.~Gschwendtner}
\affiliation{CERN, 1211 Geneva 23, Switzerland}
\author{H.~Jaworska}
\affiliation{Heinrich-Heine-Universit{\"a}t D{\"u}sseldorf, 40225 D{\"u}sseldorf, Germany}

\author{J.~Mezger}
\affiliation{Max Planck Institute for Physics, Munich, Germany}

\author{M.~Moreira}
\affiliation{CERN, 1211 Geneva 23, Switzerland}
\author{P.~Muggli}
\affiliation{CERN, 1211 Geneva 23, Switzerland}
\affiliation{Max Planck Institute for Physics, Munich, Germany}

\author{F.~Pannell}
\affiliation{University College London, London, UK}

\author{L.~Ranc}
\affiliation{Max Planck Institute for Physics, Munich, Germany}
\author{M.~Turner,}
\affiliation{CERN, 1211 Geneva 23, Switzerland}
\collaboration{AWAKE Collaboration}

\author{C.C.~Ahdida}
\affiliation{CERN, 1211 Geneva 23, Switzerland}
\author{Y. Alekajbaf}
\affiliation{Uppsala University, Uppsala, Sweden}
\author{C.~Amoedo}
\affiliation{CERN, 1211 Geneva 23, Switzerland}
\author{O.~Apsimon}
\affiliation{University of Manchester, Manchester M13 9PL, United Kingdom}
\affiliation{Cockcroft Institute, Warrington WA4 4AD, United Kingdom}
\author{R.~Apsimon}
\affiliation{Cockcroft Institute, Warrington WA4 4AD, United Kingdom}
\affiliation{Lancaster University, Lancaster LA1 4YB, United Kingdom}
\author{T.~Bachmann}
\affiliation{CERN, 1211 Geneva 23, Switzerland}
\author{C.~Badiali}
\affiliation{GoLP/Instituto de Plasmas e Fus\~{a}o Nuclear, Instituto Superior T\'{e}cnico, Universidade de Lisboa, 1049-001 Lisbon, Portugal}
\author{M.~Baquero}
\affiliation{Ecole Polytechnique Federale de Lausanne (EPFL), Swiss Plasma Center (SPC), 1015 Lausanne, Switzerland}
\author{A.~Boccardi}
\affiliation{CERN, 1211 Geneva 23, Switzerland}
\author{T.~Bogey}
\affiliation{CERN, 1211 Geneva 23, Switzerland}
\author{S.~Burger}
\affiliation{CERN, 1211 Geneva 23, Switzerland}
\author{P.N.~Burrows}
\affiliation{John Adams Institute for Accelerator Science, University of Oxford, Oxford, UK}
\author{B.~Buttensch{\"o}n}
\affiliation{Max Planck Institute for Plasma Physics, 17491 Greifswald, Germany}
\author{A.~Caldwell}
\affiliation{Max Planck Institute for Physics, Munich, Germany}
\author{M.~Chung}
\affiliation{POSTECH, Pohang 37673, Republic of Korea}
\author{C.C.~Cobo}
\affiliation{Imperial College London, London SW7 2AZ, United Kingdom}
\author{D.A.~Cooke}
\affiliation{University College London, London, UK}
\author{D.~Dancila}
\affiliation{Uppsala University, Uppsala, Sweden}
\author{C.~Davut}
\affiliation{University of Manchester, Manchester M13 9PL, United Kingdom}
\affiliation{Cockcroft Institute, Warrington WA4 4AD, United Kingdom}
\author{G.~Demeter}
\affiliation{HUN-REN Wigner Research Centre for Physics, Budapest, Hungary}
\author{A.C.~Dexter}
\affiliation{Cockcroft Institute, Warrington WA4 4AD, United Kingdom}
\affiliation{Lancaster University, Lancaster LA1 4YB, United Kingdom}
\author{S.~Doebert}
\affiliation{CERN, 1211 Geneva 23, Switzerland}
\author{A.~Eager}
\affiliation{CERN, 1211 Geneva 23, Switzerland}
\author{D.~Easton}
\affiliation{GWA, Cambridge, CB4 0WS UK}
\author{B.~Elward}
\affiliation{University of Wisconsin, Madison, WI 53706, USA}
\author{J.~Farmer}
\affiliation{Max Planck Institute for Physics, Munich, Germany}
\author{R.~Fonseca}
\affiliation{DCTI/ISCTE, Instituto Universitario de Lisboa, 1649-026, Lisboa, Portugal}
\affiliation{GoLP/Instituto de Plasmas e Fus\~{a}o Nuclear, Instituto Superior T\'{e}cnico, Universidade de Lisboa, 1049-001 Lisbon, Portugal}
\author{I.~Furno}
\affiliation{Ecole Polytechnique Federale de Lausanne (EPFL), Swiss Plasma Center (SPC), 1015 Lausanne, Switzerland}
\author{D.~Ghosal}
\affiliation{Cockcroft Institute, Warrington WA4 4AD, United Kingdom}
\affiliation{University of Liverpool, Liverpool L69 7ZE, United Kingdom}
\author{E.~Granados}
\affiliation{CERN, 1211 Geneva 23, Switzerland}
\author{J.~Gregory}
\affiliation{Cockcroft Institute, Warrington WA4 4AD, United Kingdom}
\affiliation{Lancaster University, Lancaster LA1 4YB, United Kingdom}
\author{O.~Grulke}
\affiliation{Max Planck Institute for Plasma Physics, 17491 Greifswald, Germany}
\affiliation{Technical University of Denmark, 2800 Kgs. Lyngby, Denmark}
\author{E.~Guran}
\affiliation{CERN, 1211 Geneva 23, Switzerland}
\author{D.~Harryman}
\affiliation{CERN, 1211 Geneva 23, Switzerland}
\author{M.~Hibberd}
\affiliation{University of Manchester, Manchester M13 9PL, United Kingdom}
\affiliation{Cockcroft Institute, Warrington WA4 4AD, United Kingdom}
\author{P.~Karataev}
\affiliation{Royal Holloway University of London, Egham, Surrey, TW20 0EX, United Kingdom}
\author{R.~Karimov}
\affiliation{Ecole Polytechnique Federale de Lausanne (EPFL), Swiss Plasma Center (SPC), 1015 Lausanne, Switzerland}
\author{M.A.~Kedves}
\affiliation{HUN-REN Wigner Research Centre for Physics, Budapest, Hungary}
\author{F.~Kraus}
\affiliation{Universität Bonn, 53121 Bonn, Germany}
\author{M.~Krupa}
\affiliation{CERN, 1211 Geneva 23, Switzerland}
\author{T.~Lefevre}
\affiliation{CERN, 1211 Geneva 23, Switzerland}
\author{N.~Lopes}
\affiliation{GoLP/Instituto de Plasmas e Fus\~{a}o Nuclear, Instituto Superior T\'{e}cnico, Universidade de Lisboa, 1049-001 Lisbon, Portugal}
\author{K.~Lotov}
\noaffiliation,
\author{J.~Mcgunigal}
\affiliation{University of Manchester, Manchester M13 9PL, United Kingdom}
\affiliation{Cockcroft Institute, Warrington WA4 4AD, United Kingdom}
\author{B.~Moser}
\affiliation{CERN, 1211 Geneva 23, Switzerland}
\author{Z.~Najmudin}
\affiliation{Imperial College London, London SW7 2AZ, United Kingdom}
\author{S.~Norman}
\affiliation{University of Manchester, Manchester M13 9PL, United Kingdom}
\affiliation{Cockcroft Institute, Warrington WA4 4AD, United Kingdom}
\author{N.~Okhotnikov}
\noaffiliation
\author{A.~Omoumi}
\affiliation{CERN, 1211 Geneva 23, Switzerland}
\author{C.~Pakuza}
\affiliation{CERN, 1211 Geneva 23, Switzerland}
\author{A.~Pardons}
\affiliation{CERN, 1211 Geneva 23, Switzerland}
\author{J.~Pisani}
\affiliation{GWA, Cambridge, CB4 0WS UK}
\author{A.~Pukhov}
\affiliation{Heinrich-Heine-Universit{\"a}t D{\"u}sseldorf, 40225 D{\"u}sseldorf, Germany}
\author{R.~Rossel}
\affiliation{CERN, 1211 Geneva 23, Switzerland}
\author{H.~Saberi}
\affiliation{University of Manchester, Manchester M13 9PL, United Kingdom}
\affiliation{Cockcroft Institute, Warrington WA4 4AD, United Kingdom}
\author{M.V.~dos Santos}
\affiliation{CERN, 1211 Geneva 23, Switzerland}
\author{O.~Schmitz}
\affiliation{University of Wisconsin, Madison, WI 53706, USA}
\author{F.~Sharmin}
\affiliation{University of Wisconsin, Madison, WI 53706, USA}
\author{F.~Silva}
\affiliation{INESC-ID, Instituto Superior Técnico, Universidade de Lisboa, 1049-001 Lisbon, Portugal}
\author{L.~Silva}
\affiliation{GoLP/Instituto de Plasmas e Fus\~{a}o Nuclear, Instituto Superior T\'{e}cnico, Universidade de Lisboa, 1049-001 Lisbon, Portugal}
\author{B.~Spear}
\affiliation{John Adams Institute for Accelerator Science, University of Oxford, Oxford, UK}
\author{L.~Stant}
\affiliation{CERN, 1211 Geneva 23, Switzerland}
\author{C.~Stollberg}
\affiliation{Ecole Polytechnique Federale de Lausanne (EPFL), Swiss Plasma Center (SPC), 1015 Lausanne, Switzerland}
\author{A.~Sublet}
\affiliation{CERN, 1211 Geneva 23, Switzerland}
\author{C.~Swain}
\affiliation{Cockcroft Institute, Warrington WA4 4AD, United Kingdom}
\affiliation{University of Liverpool, Liverpool L69 7ZE, United Kingdom}
\author{G.~Tenasini}
\affiliation{CERN, 1211 Geneva 23, Switzerland}
\author{A.~Topaloudis}
\affiliation{CERN, 1211 Geneva 23, Switzerland}
\author{P.~Tuev}
\noaffiliation
\author{J.~Uncles}
\affiliation{GWA, Cambridge, CB4 0WS UK}
\author{F.~Velotti}
\affiliation{CERN, 1211 Geneva 23, Switzerland}
\author{J.~Vieira}
\affiliation{GoLP/Instituto de Plasmas e Fus\~{a}o Nuclear, Instituto Superior T\'{e}cnico, Universidade de Lisboa, 1049-001 Lisbon, Portugal}
\author{C.~Welsch}
\affiliation{Cockcroft Institute, Warrington WA4 4AD, United Kingdom}
\affiliation{University of Liverpool, Liverpool L69 7ZE, United Kingdom}
\author{T.~Wilson}
\affiliation{Heinrich-Heine-Universit{\"a}t D{\"u}sseldorf, 40225 D{\"u}sseldorf, Germany}
\author{M.~Wing}
\affiliation{University College London, London, UK}
\author{J.~Wolfenden}
\affiliation{Cockcroft Institute, Warrington WA4 4AD, United Kingdom}
\affiliation{University of Liverpool, Liverpool L69 7ZE, United Kingdom}
\author{B.~Woolley}
\affiliation{CERN, 1211 Geneva 23, Switzerland}
\author{G.~Xia}
\affiliation{Cockcroft Institute, Warrington WA4 4AD, United Kingdom}
\affiliation{University of Manchester, Manchester M13 9PL, United Kingdom}
\author{V.~Yarygova}
\noaffiliation
\author{W.~Zhang}
\affiliation{John Adams Institute for Accelerator Science, University of Oxford, Oxford, UK}

\date{\today}

\sloppy
\begin{abstract}

This manuscript proposes a method to enable controlled high-gradient particle acceleration when requiring self-modulation of the drive bunch. While electron bunch seeding of self-modulation (eSSM) has been realised at a plasma electron density \unit[$n_\mathrm{pe}\cong10^{14}$]{cm$^{-3}$}~\cite{prllv}, it has not been demonstrated at higher plasma densities due to limitations of available seed bunch properties.
As experimentally shown in this manuscript, truncating available seed bunches with a relativistic ionisation front allows these limitations to be overcome. This seeding method is called truncated electron bunch seeding of self-modulation (teSSM) and experiments confirm that --when using teSSM -- self-modulation becomes reproducible at \unit[$n_\mathrm{pe}=7\times10^{14}$]{cm$^{-3}$}. Additionally, the seed wakefield amplitude is also increased, which is known to be advantageous because it shortens the length needed to reach self-modulation saturation.  The presented results establish teSSM as a method for achieving controlled, high-gradient particle acceleration with long drivers and available seed bunches.\\

\end{abstract}

\maketitle
\null
\clearpage

Plasma wakefields offer a promising route for high-gradient acceleration of charged particles. When using plasma electron densities \unit[$n_\mathrm{pe}\gtrsim10^{14}$]{cm$^{-3}$},
accelerating fields \unit[$\gtrsim$]{GV/m} can be reached~\cite{Esarey2009RMP,Tajima1979PRL}.
Together with energetic (\unit[$\gtrsim10$]{kJ}) relativistic drivers, energy gains \unit[$\gtrsim100$]{GeV} can be achieved in a single plasma stage~\cite{Caldwell2009NatPhys,nature}.

However, a challenge remains: at present, such available energetic drivers are relatively long, and therefore, of relatively low density, $n_\mathrm{b} < n_\mathrm{pe}$. This is because their duration ($\sigma_{\mathrm{t,d}}$) exceeds the characteristic time-scale of the wakefields given by 
$\tau_\mathrm{pe}$, where $\tau_\mathrm{pe} = {2\pi}/{\omega_\mathrm{pe}}$.
Here, $\omega_\mathrm{pe} = \sqrt{{n_\mathrm{pe} e^2}/{m_\mathrm{e} \varepsilon_0}}$ is the plasma electron angular frequency, where $m_\mathrm{e}$ and $e$ are the electron mass and charge, respectively, and $\varepsilon_0$ is the vacuum permittivity.

Generally, long and low-density bunches cannot effectively drive high amplitude wakefields. Even when the bunch has a sharp onset, the low $n_\mathrm{b}$ always results in low field amplitude $E_z$, which can be estimated by
\begin{equation}
\begin{aligned}
E_z &\sim \frac{n_\mathrm{b}}{n_\mathrm{pe}}\,E_\mathrm{wb},
\end{aligned}\label{Eq:seed_wakefields}
\end{equation}
where $E_\mathrm{wb}~=~{m_\mathrm{e}c\omega_\mathrm{pe}}/{e}$ is the cold plasma wavebreaking field and $c$ is the vacuum speed of light.
For example, for typical experimental values of \unit[$n_\mathrm{b}=8\times10^{12}$]{cm$^{-3}$} and \unit[$n_\mathrm{pe}=7~\times~10^{14}$]~{cm$^{-3}$}~\cite{symmetry}, \unit[$E_z~\sim~30$]{MV/m}, far below the possible GV/m when $n_\mathrm{b}\sim n_\mathrm{pe}$.

To drive \unit[$\gtrsim$]{GV/m} wakefields, initially long and low-density drivers must be transformed into trains of micro-bunches, with micro-bunches of periodicity $\tau_\mathrm{pe}$.
Such micro-bunch trains can resonantly drive wakefields, in which case 
\begin{equation}
    \label{Eq:seednb}
    E_z~\sim~N_{mb}\frac{n_\mathrm{b}}{n_\mathrm{pe}} E_\mathrm{wb},
\end{equation} where $N_{mb}$ is the number of micro-bunches.
For example, using the same $n_\mathrm{b}$ and $n_\mathrm{pe}$ as above, and assuming $N_{mb}~=~50$,  \unit[$E_z\sim1.5$]{GV/m}.

One way of transforming initially long drivers into trains of micro-bunches is the self-modulation (SM) process~\cite{seedMT2, dense, growth}: drivers entering plasma excite transverse and longitudinal wakefields with amplitudes equal or lower than those given by Eq.~\ref{Eq:seed_wakefields}. When $\sigma_{\mathrm{t,d}}\gg\tau_\mathrm{pe}$, the transverse wakefields act periodically focusing and defocusing on the initially long drivers, longitudinally modulating their $n_\mathrm{b}(t)$. Here the time coordinate $t$ corresponds to the longitudinal position along the driver, in the co-moving frame (moving at $c$). The effect is self-reinforcing, whereby the increasingly modulated drivers excite higher amplitude wakefields, which in turn increases their density modulation, until the micro-bunch trains are fully formed~\cite{dense, short}.
SM can grow from noise~\cite{SM}, e.g. from random perturbations in the density profiles of the drivers. In this case, the process is called the self-modulation instability (SMI), and the phase and amplitude of the wakefields are inherently random.
This poses a challenge when accelerating charged particles (i.e. witness bunches), which must always be injected at a precise wakefield phase (time).
\vspace{0.2cm}

\begin{figure}[htb!]
\centering
    \includegraphics[width=\linewidth]{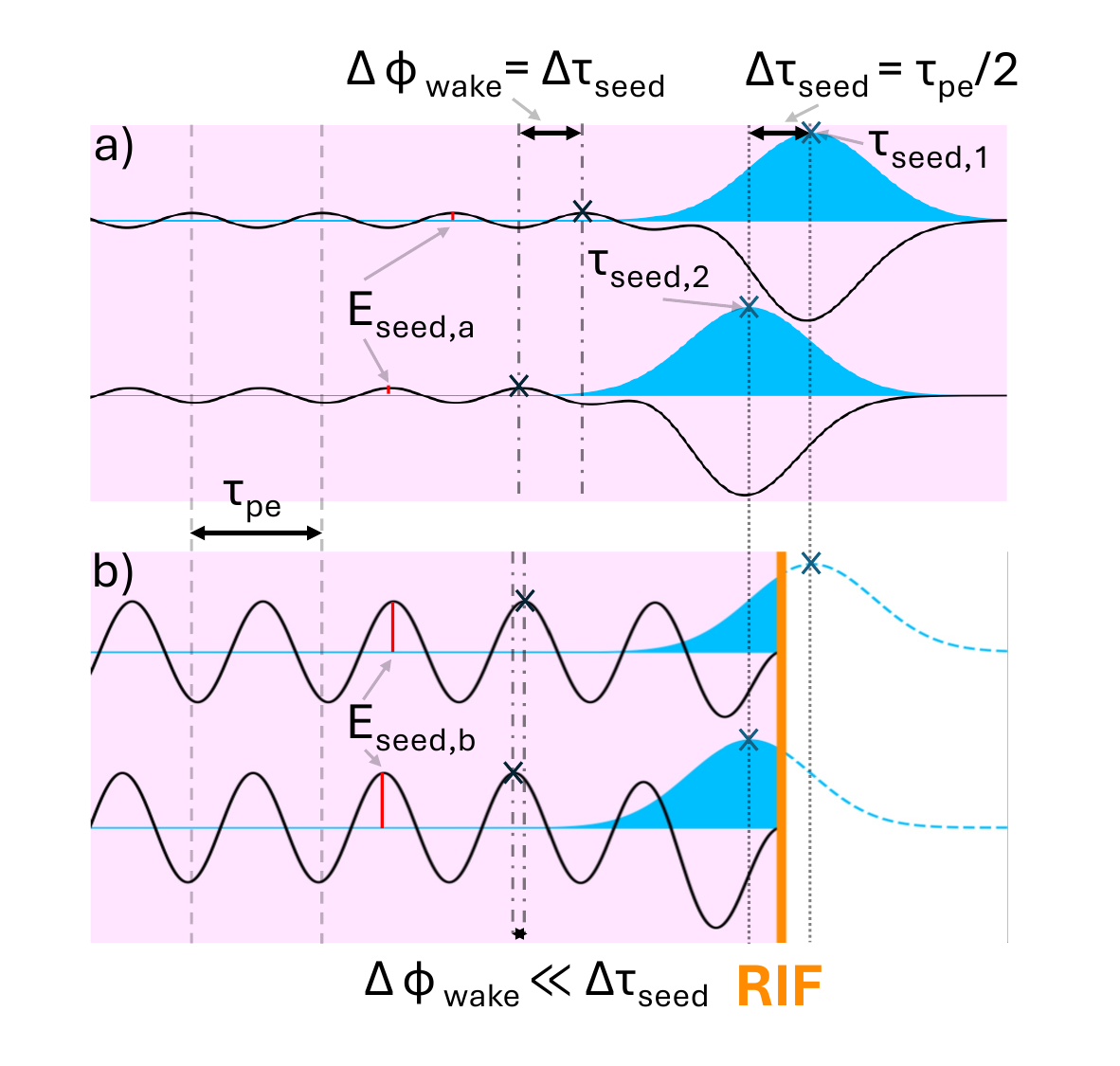}
    \caption{Transverse wakefields (black solid lines) in plasma (pink shaded areas), driven by Gaussian seed pulses (blue shaded areas). Two vertical gray dashed lines mark the wakefield period $\tau_\mathrm{pe}$. Figures a and b: wakefield phase difference $\Delta\phi_{\mathrm{wake}}$ (difference between wakefield peak locations, marked by the black crosses and the two vertical gray dotted-dashed lines) from seed arrival-time jitter $\Delta\tau_{\mathrm{seed}}$ (difference between $\tau_\mathrm{seed,1}$ and $\tau_\mathrm{seed,2}$, marked by the blue crosses and the two vertical gray dotted lines) for the eSSM (a) and teSSM (b). Field amplitudes $E_{\textrm{seed}}$ (vertical red solid lines) on the same vertical scale. Figure b: solid vertical orange line is the RIF position. Bunches and wakefields propagate to the right.}
    \label{fig:illustrationteSSM}
\end{figure}

It has been demonstrated experimentally~\cite{prllv, IFR} that phase reproducibility can be achieved by seeding SM, i.e., when providing the same initial perturbation (seed) from which SM grows. When seeded, SM is referred to as seeded self-modulation (SSM). There are three requirements on the seed for SSM to occur:
\begin{enumerate}
    \item \textbf{Dephasing}: the relativistic gamma factor of the seed ($\gamma_{\textrm{seed}}$), should be approximately equal to that of the self-modulating driver $\gamma_{SM}$. When they differ, the seed and the driver dephase by a distance
    \begin{equation}\label{dephase}
        \Delta L \simeq \frac{1}{2}\left(\frac{1}{\gamma_{SM}^2}-\frac{1}{\gamma_{\textrm{seed}}^2}\right) L_\mathrm{seed},
    \end{equation}
    where $L_\mathrm{seed}$ is the seeding distance, i.e. the distance over which the seed must interact with the driver to initiate SSM.  Both $\gamma_{SM}$ and $\gamma_{\textrm{seed}}$ vary along the plasma~\cite{short, Lotov2015_PhysicsSM, Pukhov}. Successful seeding requires $\Delta L \ll c\tau_\mathrm{pe}$. 
    
    Previous experimental results~\cite{prllv} show SSM for $\Delta L/L_\mathrm{seed}\lesssim c\tau_\mathrm{pe}/8$ (with estimated \unit[$L_\mathrm{seed}\sim2$]{m}), or when expressed as duration $\Delta\tau/\tau_\mathrm{pe}\lesssim\tau_\mathrm{pe}/8$, where $\Delta\tau$ is the accumulated time difference between the seed and driver over $L_\mathrm{seed}$. 
    \item \textbf{Amplitude:} the amplitude of the wakefields driven by the seed $E_{\textrm{seed}}$ must greatly exceed the amplitude of the noise $E_{\textrm{noise}}$~\cite{SM}:
    \begin{equation}
        E_{\textrm{seed}}\gg E_{\textrm{noise}}.
    \end{equation}
    Greater $E_{\textrm{seed}}$ shortens the length of plasma required to complete SM, also called saturation~\cite{Saturationlength}.
    \item \textbf{Reproducibility:} as illustrated in Fig.~\ref{fig:illustrationteSSM}a using  linear wakefield theory calculations~\cite{Keinigs}, variations ($\Delta\tau_\mathrm{seed}$) in seed arrival time $\tau_\mathrm{seed}$ or onset translate into identical variations of wakefield phase $\Delta\phi_{\mathrm{wake}}$~\cite{Pukhov}, therefore requiring: 
    \begin{equation}
        (\Delta\tau_\mathrm{seed}\simeq\Delta\phi_{\mathrm{wake}})\ll \tau_\mathrm{pe}.
    \end{equation}
\end{enumerate}

\vspace{0.2cm}

One method of seeding SM is with a relativistic ionisation front (RIF)~\cite{IFR}, which creates plasma from gas. For RIF seeding, each short laser pulse propagates through gas, co-linearly and temporally overlapped with the driver. The transition to plasma from gas occurs sharply at the leading edge of the laser pulse~\cite{IFR, laser, laser2}.
The plasma experiences a sudden onset in drive bunch density at the RIF, which determines the phase and amplitude of the wakefields~\cite{IFR}.

RIF seeding was experimentally demonstrated using \unit[400]{GeV} proton drivers ($\gamma_{SM}=427$) in plasma with \unit[$n_\mathrm{pe}\approx1-10\times 10^{14}$]{cm$^{-3}$} and laser pulses of central wavelength \unit[$\lambda_l=780$]{nm} ($\gamma_{\textrm{seed}}\approx 1600$) (meeting condition \textbf{1.\,Dephasing}), with a seed amplitude \unit[$>4$]{MV/m}\,$\gg\,$\unit[10]{kV/m} (meeting condition \textbf{2.\,Amplitude})~\cite{SM,IFR} and $\Delta\tau_\mathrm{seed}\approx~$\unit[0.1]{ps}$~\ll\tau_\mathrm{pe}$\unit[$\sim11-3.5$]{ps} (meeting condition \textbf{3.\,Reproducibility})~\cite{laser, laser2}.

RIF seeding has several limitations: the part of the drivers located ahead of the RIF remains in gas and therefore does not contribute to the wakefields; this part could undergo SMI when entering a subsequent pre-ionised plasma (e.g. second plasma stage), potentially disrupting the wakefields~\cite{Lotov2015_PhysicsSM, liviofilamentation}. Furthermore, the seed wakefield amplitude cannot be adjusted independently of the driver parameters and of the RIF position.

\begin{figure*}[htb!]
\centering
    {\includegraphics[width = 1\textwidth]{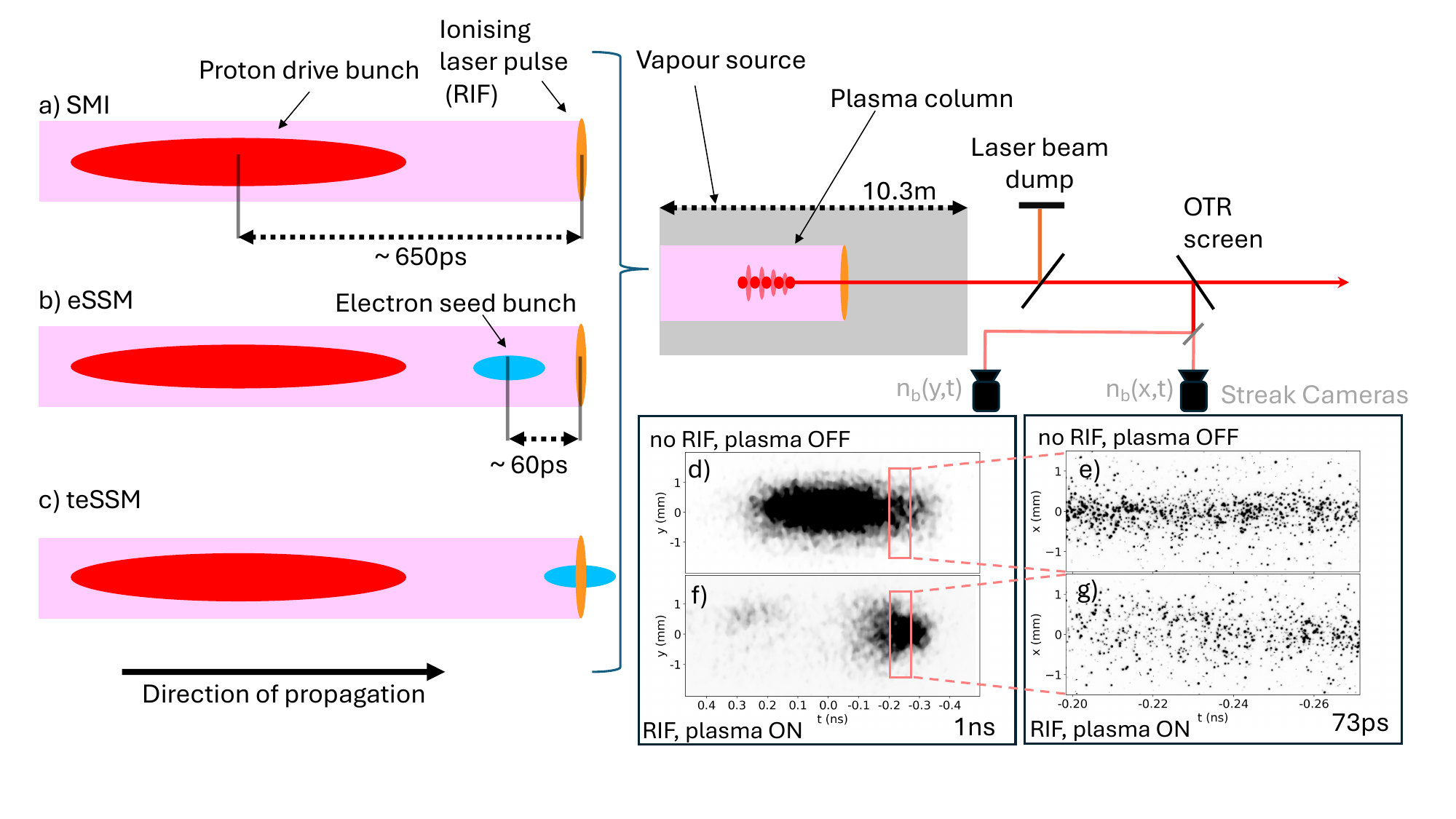}} 
        \caption{
\label{Fig2}
Schematic of the experimental setup. Proton drivers (red) propagate through plasma (pink) in the vapour source (grey). Rubidium vapour ionised by the RIF (orange). There are three configurations: SMI (a): drivers preceded by a RIF; eSSM (b): drivers preceded by seed bunches, both preceded by a RIF; teSSM (c): drivers preceded by seed bunches, which are overlapped in space and time with the RIF. Downstream the plasma and the laser beam dump, longitudinal proton bunch densities are measured using an OTR screen and two streak cameras~\cite{karlrieger}. One streak camera measures the time resolved drive bunch density with a time window of \unit[1]{ns} ($n_\mathrm{b}(y,t)$) and the second one with a time window of \unit[73]{ps} ($n_\mathrm{b}(x,t)$), where $x$ and $y$ are perpendicular transverse dimensions and $t$ is axis along the bunch. Sub-figures d and e: measurements without plasma, in vacuum; f and g: and with plasma (in configuration a). Bunches and pulses propagate to the right.}

\end{figure*}

Alternatively, wakefields from preceding bunches
can seed SM, enabling modulation of the entire driver and independent adjustment of the seed wakefield amplitude~\cite{prllv}. This was experimentally demonstrated using \unit[$\sim 20$]{MeV} electron bunches ($\gamma_{\textrm{seed}}\sim40$) in plasma (eSSM) with \unit[$n_\mathrm{pe}\approx1\times 10^{14}$]{cm$^{-3}$}~\cite{prllv}.

\vspace{0.2cm}

Meeting all three SSM conditions becomes increasingly difficult at higher $n_\mathrm{pe}$, because $\tau_\mathrm{pe}$ decreases and conditions \textbf{1.\,Dephasing} and \textbf{3.\,Reproducibility} become more difficult to meet. An example of typically available bunch and plasma parameters that are not compatible can be found in the~\textit{Appendix}. A higher $n_\mathrm{pe}$ is generally preferred (e.g. \unit[$n_\mathrm{pe}=7\times 10^{14}$]{cm$^{-3}$} for AWAKE), since it enables higher accelerating fields with appropriate drivers (Eqs.~\ref{Eq:seed_wakefields} and~\ref{Eq:seednb} with \unit[$E_\mathrm{wb}\sim2.6$]{GV/m}). At such $n_\mathrm{pe}$, eSSM does not work with available seed bunches (e.g. see~\textit{Appendix~\ref{app1}}). However, there is a solution: truncating electron seed bunches with a RIF. This can be realised by overlapping every seed bunch with a RIF (as illustrated in Fig.~\ref{fig:illustrationteSSM}b), creating a sharp plasma onset within the seed. The SSM conditions are then as follows:

\begin{enumerate}
    \item \textbf{Dephasing}: $\gamma_{\textrm{seed}}$ is now defined by the RIF and becomes $\gamma_{\textrm{RIF}}$, provided the RIF is inside the seed bunch. When maximising seed bunch charge in plasma (optimal in experiments), the optimal initial starting placement for the RIF inside of the seed (derived from Eq.~\ref{dephase}) is given by:
    \begin{equation}\label{optimal}
        t_\mathrm{RIF}^\mathrm{opt} = 2\sigma_\mathrm{t,s} + \frac{L_\mathrm{seed}}{2} \left(\frac{1}{\gamma_\mathrm{RIF}^2} - \frac{1}{\gamma_{\textrm{bunch}}^2}\right), 
    \end{equation} 
    where $\gamma_\mathrm{bunch}$ is the relativistic gamma factor of the seed bunch and $\sigma_\mathrm{t,s}$ its duration. Since $\gamma_{\mathrm{RIF}} > \gamma_\mathrm{bunch}$, an increasing fraction of the seed bunch charge is in plasma and experiences focusing as the bunches propagate. This is illustrated by linear theory calculations in Fig.~\ref{fig:illustrationteSSM}b, where the seed bunch moves later by $\tau_{\mathrm{pe}}/2$ with respect to the RIF and the additional seed charge that enters the plasma induces only a small wakefield phase difference $\Delta\phi_{\mathrm{wake}}\sim\tau_{\mathrm{pe}}/8\ll\Delta\tau_{\mathrm{seed}}$.
    \item \textbf{Amplitude:} when seed bunches are longer than optimal ($\sigma_{t,s}>\tau_{\mathrm{pe}}/(\sqrt{2}\pi)$), they can be shortened by the RIF to increase $E_{\mathrm{seed}}$~\cite{Keinigs}, as illustrated in Figs.~\ref{fig:illustrationteSSM}a and b by comparing the field amplitudes behind the seeds (red vertical lines), which show $E_{\mathrm{seed},b}\simeq 6\times E_{\mathrm{seed,a}}$.
    \item \textbf{Reproducibility:} differences in the seed arrival time $(\Delta\tau_\mathrm{seed})$ are also illustrated by Fig.~\ref{fig:illustrationteSSM}b. The benefit is that, instead of variations in $\Delta\tau_\mathrm{seed}$ directly translating into variations in the wakefield phase $\Delta\phi_{\mathrm{wake}}$ is reduced. In the example of Fig.~\ref{fig:illustrationteSSM}b, $\Delta\phi_{\mathrm{wake}}$ is $\approx\Delta\tau_\mathrm{seed}/8$ with teSSM instead of $\approx~\Delta\tau_\mathrm{seed}$ with eSSM (Fig.~\ref{fig:illustrationteSSM}a).
\end{enumerate}

Experiments are conducted in AWAKE (Advanced WAKefield Experiment)~\cite{symmetry} at CERN. A schematic of the experimental setup is shown in Fig.~\ref{Fig2}, at the centre of which a \unit[10.3]{m} long rubidium vapour source (grey) provides vapour densities \unit[$n_{\mathrm{vapour}}=1-10\times10^{14}$]{cm$^{-3}$}~\cite{plasmacell}.
The rubidium atoms are singly ionised by a \unit[($100\pm5$)]{mJ}, \unit[120]{fs} laser pulse, 
creating a plasma column (pink) with density equal to that of the rubidium vapour~\cite{density}. In this manuscript, measurements are performed with \unit[$n_\mathrm{pe}=(7.01\pm 0.05)\times10^{14}$]{cm$^{-3}$} (AWAKE density baseline for acceleration).

The drivers are \unit[400]{GeV} proton bunches provided by the CERN SPS of charge \unit[($46000\pm2000$)]{pC}. They are longitudinally approximately Gaussian with an RMS duration of \unit[$\sigma_\mathrm{t,p}=(170\pm8$)]{ps}~$\gg \tau_\mathrm{pe}$, and RMS transverse sizes at the entrance of the plasma of \unit[$(200\pm15$)]{$\mu$m}.
Their RMS arrival time jitter is \unit[$(5.8\pm0.4)$]{ps}.

Electron seed bunches (with an energy per particle of \unit[17.8]{MeV}) are produced by an S-band RF photo-injector~\cite{rf} and transported to the entrance of the vapour source by a \unit[15]{m} long transfer line~\cite{injector, beamline}. The last quadrupole triplet in the transfer line focuses the seed bunches and places the waist (RMS transverse sizes \unit[$\sigma_\mathrm{r,s}=(200~\pm~$40)]{$\mu$m}) at the entrance of the vapour source~\cite{IPAC}. The RMS longitudinal duration of the seed bunches is \unit[$\sigma_\mathrm{t,s}=(2.1\pm0.5$)]{ps} for a charge per bunch of \unit[$(200\pm20$)]{pC}. Due to diagnostics limitations, the arrival time jitter cannot be quantified.

Each ionising laser pulse in vapour creates a RIF. For each event and measurement configuration (Fig.~\ref{Fig2}a--c) the RIF is positioned \unit[$t=+~650$]{ps} ahead of the driver centre, which is defined as $t=0$. However, the seeding method differs in each configuration. For SMI (Fig.~\ref{Fig2}a) there is no seed bunch, the RIF arrives much before the driver and SM starts from driver noise. For eSSM (Fig.~\ref{Fig2}b), the electron bunch seed arrives \unit[($60\pm1$)]{ps} after the RIF and is therefore entirely in plasma. For teSSM (Fig.~\ref{Fig2}c) the electron bunch is positioned such that it overlaps with the RIF at $t_\mathrm{RIF}^\mathrm{opt}\sim+\sigma_\mathrm{t,s}$ (optimum from Eq.~\ref{optimal} with $\gamma_\mathrm{s}\sim 40$, $\gamma_\mathrm{RIF}\sim 1600$ and an estimated \unit[$L_\mathrm{seed}=2$]{m}).

Downstream of the plasma, proton drivers traverse an aluminium-coated metallic screen, where they emit optical transition radiation (OTR). The backward OTR is split by a 45:55 beamsplitter, filtered with a bandpass filter transmitting only wavelengths within a range of \unit[($450\pm25$)]{nm}, and imaged onto the entrance slits of two streak cameras. Each slit selects a \unit[$\Delta x, \Delta y\sim80$]{$\mu$m}-wide region around the bunch axis, with the two slits oriented orthogonally in the $x$ and $y$ planes~\cite{streak_camera}. Time-resolved driver bunch densities, $n_\mathrm{b}(x,t)$ and $n_\mathrm{b}(y,t)$, are thus measured simultaneously over time windows of \unit[73]{ps} and \unit[1]{ns}, respectively. Typical measurements of the bunch after propagation in vacuum (no plasma) are shown in Figs.~\ref{Fig2}d and e. In this case, the bunch structure is approximately Gaussian on the \unit[1]{ns} timescale and approximately uniform on the \unit[73]{ps} timescale, as expected when the driver duration ($\sigma_{\mathrm{t,p}} \approx$ \unit[170]{ps}) is much longer than the measurement time window (\unit[73]{ps}). 

\begin{figure}[htb!]
    \centering    \includegraphics[width=0.5\textwidth]{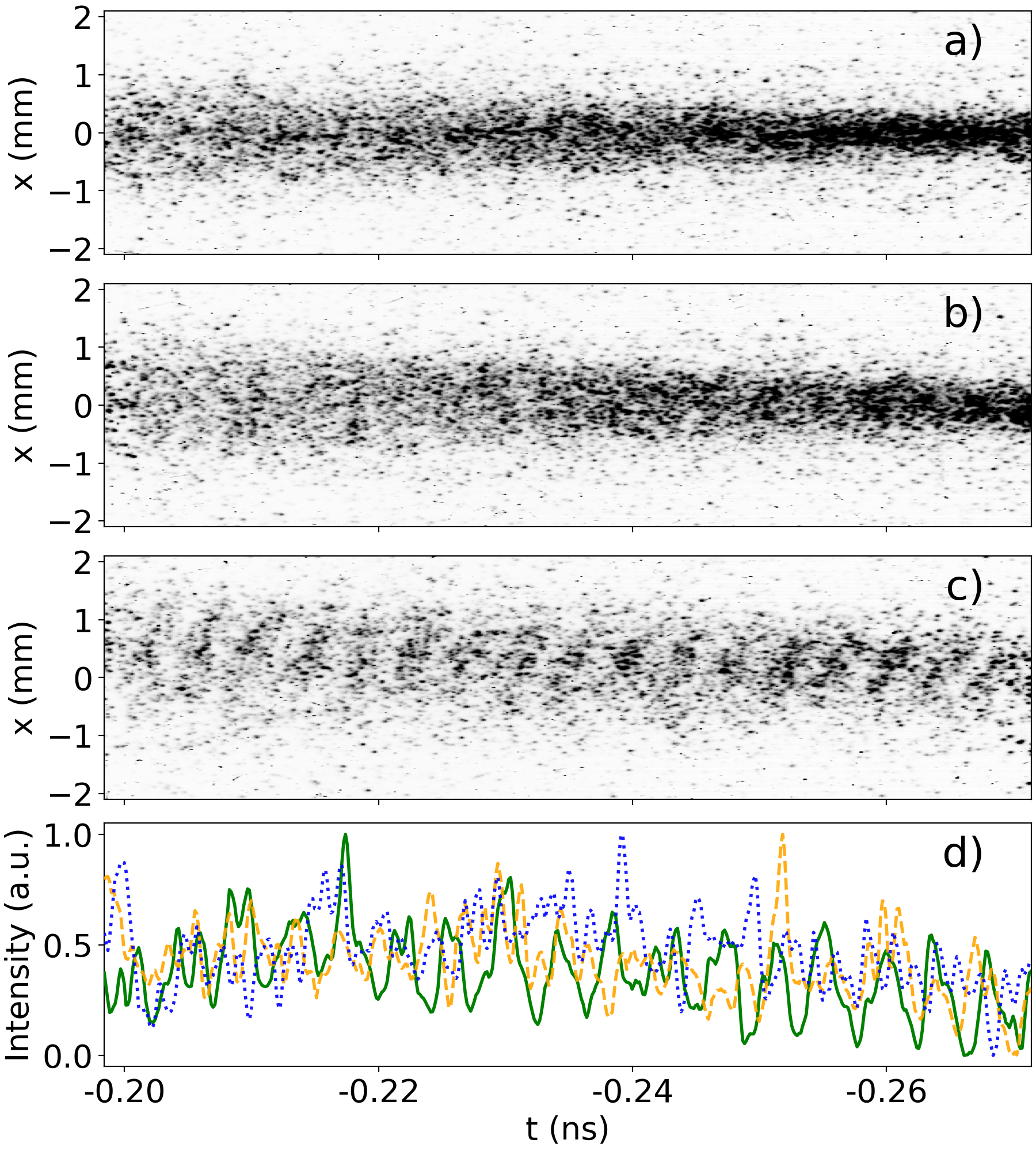}
    \caption{
        \label{fig:microbunches} 
       Measurements of twelve summed proton bunch density profiles $n_\mathrm{b}(x,t)$, aligned in time following the same procedure as in Ref.~\cite{IFR}, when using SMI (a), eSSM (b), and teSSM (c). Figures a-c: on the same colour scale and identical streak camera settings. Figure d: vertical projections for SMI (a, blue dotted line), eSSM (b, orange dashed line) and teSSM (c, green solid line). Intensity normalised to the peak value. Bunches propagate to the right.} 
\end{figure}

Measurements of $n_\mathrm{b}(x,t)$ and $n_\mathrm{b}(y,t)$ are then also performed in the presence of plasma (Figs.~\ref{Fig2}f and g). In this case, the measured $n_\mathrm{b}(y,t)$ (Fig.~\ref{Fig2}f) differs significantly from those obtained without plasma, as expected when SM develops~\cite{density,growth}. Similarly, $n_\mathrm{b}(x,t)$ measurements with plasma (Fig.~\ref{Fig2}g) show subtle differences, particularly a lower bunch density due to proton defocusing by the wakefields. However, due to measurement noise and the limited temporal resolution of the streak camera, micro-bunches at this $n_{\mathrm{pe}}$ are not well defined for all three scenarios. Unfortunately, the measurement quality is insufficient to determine the micro-bunch timing and frequency from all individual measurements, and therefore to distinguish between SSM and SMI.

However, the key feature of SSM is that the micro-bunch timing is reproducible from event to event, which can still be assessed by summing multiple measurements of $n_\mathrm{b}(x,t)$, as shown in Fig.~\ref{fig:microbunches}. Summing is made possible by using an absolute time reference with respect to the RIF, as described in Ref.~\cite{IFR}. When the micro-bunch timing is different from event to event, the periodic longitudinal modulation averages out in the sum. This means that although SM occurs, it is not reproducible. This was observed when using either no seed (SMI, Fig.~\ref{fig:microbunches}a) or the entire electron bunch as a seed (eSSM, Fig.~\ref{fig:microbunches}b), in which case the vertical projections (blue dotted and orange dashed lines in Fig.~\ref{fig:microbunches}d) do not show any periodic structure at this $n_\mathrm{pe}$. 

Conversely, when the micro-bunch timing is reproducible —meaning that the micro-bunch positions along $t$ are consistent from event to event— the signal-to-noise ratio and therefore modulation depth increases when multiple measurements are summed, making the longitudinal structure observable.  This was the case with teSSM, where a periodic modulation (micro-bunch structure) becomes visible in the summed measurement (Fig.~\ref{fig:microbunches}c) as well as its vertical projection (green solid line in Fig.~\ref{fig:microbunches}d). The teSSM measurements in Figs.~\ref{fig:microbunches}c and d show approximately \unit[17]{micro-bunches} within the \unit[73]{ps} time window. This indicates that the observed periodic structure results from the driver undergoing SM, because the measured modulation period, $\tau = \unit[73]{ps}/17 \approx \unit[4.3]{ps}$, is close to the plasma period predicted by wakefield theory for this $n_\mathrm{pe}$, $\tau_\mathrm{pe} = 2\pi/\omega_\mathrm{pe} = \unit[(4.21\pm0.2)]{ps}$.

\begin{figure}[htb!]
\centering
\includegraphics[width=0.47\textwidth]{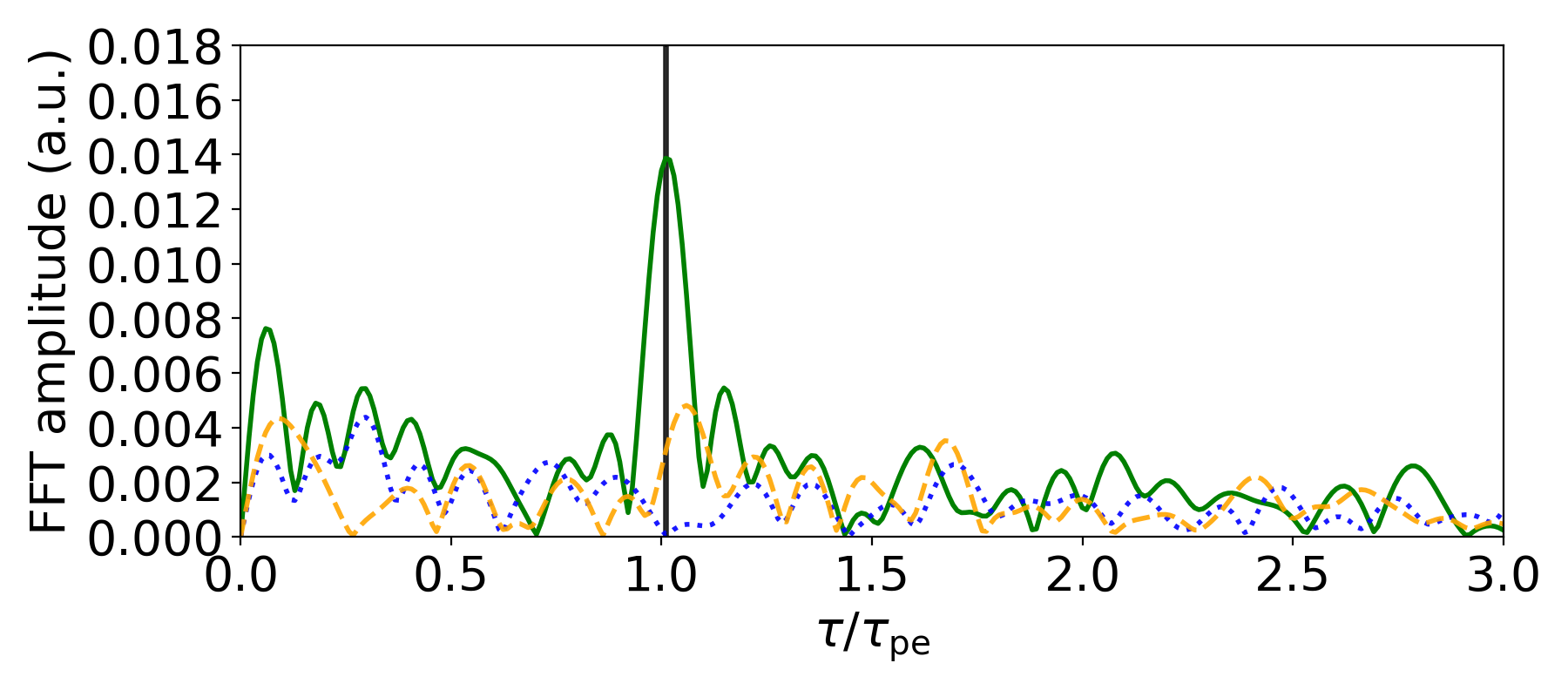}
\caption{\label{fig:frequency} Normalised amplitude resulting from the fast fourier transform (FFT) analysis as a function of the normalised period $\tau/\tau_\mathrm{pe}$ on the projections in Fig.~\ref{fig:microbunches}d, for SMI (blue dotted line), eSSM (orange dashed line) and teSSM (green solid line). Black vertical line centres around the FFT peak and its width indicates the FFT bin width.}
\end{figure}

To obtain a more accurate measurement of the wakefield period and to also test for the presence of periodic modulation in the SMI (Fig.~\ref{fig:microbunches}a) and eSSM (Fig.~\ref{fig:microbunches}b) measurements, a fast Fourier transform (FFT) analysis is performed on all the lineouts shown in Fig.~\ref{fig:microbunches}d. Results are shown in Fig.~\ref{fig:frequency}, where for SMI (blue dotted line) and eSSM (orange dashed line), the FFT frequency spectrum shows no prominent peak (max. peak amplitude: $0.0042$ (SMI), $0.0048$ (eSSM) $\ll0.014$ (teSSM)), indicating that no dominant frequency is present. This is as expected when the longitudinal micro-bunch structure is not reproducible and is thus averaged out by the summation. However, for teSSM (green solid line), the FFT frequency spectrum shows a prominent peak at the expected period, $\tau / \tau_\mathrm{pe}=1$ (indicated by the vertical black solid line). This is only possible when SM is seeded, and implies that all conditions discussed above are fulfilled.

\begin{figure}[htb!]
    \centering
    \includegraphics[width=0.5\textwidth]{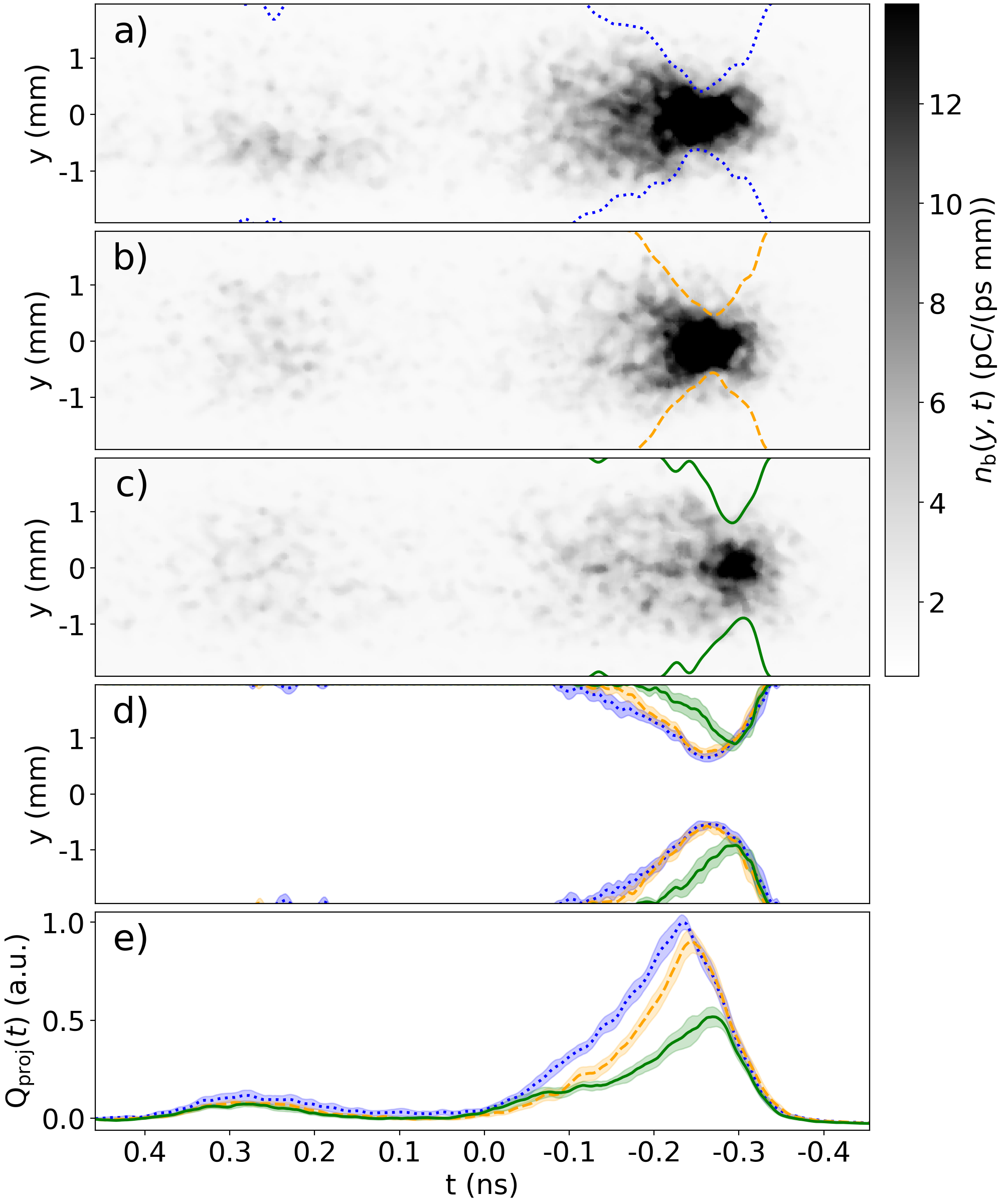}
\caption{Single measurements of $n_\mathrm{b}(y,t)$ for SMI (a), eSSM (b) and teSSM (c). Figures a-c: on the same colour scale and identical streak camera settings. Median filter~\cite{numpy} with size$=3$ applied to reduce measurement noise. Where \unit[$<\pm2$]{mm}, coloured lines in a--c mark where $n_\mathrm{b}(y)$ profiles reach \unit[20]{\%} of their peak value, as in~\cite{prllv}. Figure~d: same quantity, but displays the mean (line) and standard deviation (colorband) of the twelve measurements of Figs.~\ref{fig:microbunches}a-c for SMI (blue dotted line), eSSM (orange dashed line) and teSSM (green solid line). Figure e shows the mean (line) and standard deviations (colorband) of the vertical projections of $n_\mathrm{b}(y)$ for the twelve measurements. Bunches propagating to the right.}
\label{fig:nanosecond_streak}
\end{figure}

Simultaneously with the measurements shown in Fig.~\ref{fig:microbunches}, measurements on the nanosecond timescale in the perpendicular plane $(n_\mathrm{b}(y,t))$ are also performed. Representative single-event measurements are shown in Fig.~\ref{fig:nanosecond_streak}: for SMI (a), for eSSM (b), and for teSSM (c). 

First, the clear difference between these measurements and the no plasma case shown in Fig.~\ref{Fig2}d indicates that SM develops for all configurations and events. The indication that SM developed is the charge disappearance (\unit[$t>-0.2$]{ns}) along the bunch. Particles that are focused and defocused by the wakefields during SM diverge downstream of the plasma and their density within the streak camera slit decreases. This implies that individual measurements contributing to Figs.~\ref{fig:microbunches}a and b are from self-modulated drivers. Therefore SM must start from noise for SMI, and from either noise or the electron seed bunch for eSSM. 

However, Figs.~\ref{fig:nanosecond_streak}a (SMI) and b (eSSM) also differ. This is clear when e.g. comparing the vertical projections and their uncertainties, shown in Fig.~\ref{fig:nanosecond_streak}e: for eSSM (orange dashed line) the projection peaks at slightly earlier $t$ and at lower amplitudes (\unit[$0.90\pm0.05$]) than for SMI (blue dotted line) at \unit[$1.00\pm0.03$]. Because SM develops in the eSSM configuration and differs from SMI for the given bunch and plasma parameters (Fig.~\ref{fig:nanosecond_streak}e), the conditions for \textbf{1.\,Dephasing} and \textbf{2.\,Amplitude} must have been satisfied; otherwise, SM could not have developed. However, the lack of reproducibility in micro-bunch timing (Fig.~\ref{fig:microbunches}) shows that the \textbf{3.\,Reproducibility} condition was not met.

In addition to the measured $n_\mathrm{b}(y,t)$, Figs.~\ref{fig:nanosecond_streak}a--c (coloured lines) show the transverse sizes of each pixel-line of the bunch $n_\mathrm{b}(y)$, obtained as in Ref.~\cite{prllv}. All of these sizes, along with their uncertainties (given by the standard deviation over twelve measurements), are summarised in Fig.~\ref{fig:nanosecond_streak}d. Both the projections (Fig.~\ref{fig:nanosecond_streak}e) and the transverse bunch sizes along $t$ (Fig.~\ref{fig:nanosecond_streak}d) exhibit faster variations for teSSM than for eSSM, and both vary faster than for SMI. This indirectly gives information about the seed field amplitude $E_{\textrm{seed}}$ because higher $E_{\textrm{seed}}$ results in faster SM growth along both the plasma and the bunch. Along the bunch, this appears as a more rapid rise and fall in the vertical projections, and a faster evolution of the transverse size, as discussed in Refs.~\cite{streak_camera,ionmotion}. 

To be quantitative, one can compare e.g. at which $t$ transverse sizes reach a value of \unit[$\pm2$]{mm} in Fig.~\ref{fig:nanosecond_streak}d: before the charge peak (\unit[$t<-0.25$]{ps}) this size is reached around \unit[$t=(-0.35\pm0.05)$]{ps} for all configurations, showing that SM starts at the same $t$. However after the peak (\unit[$t>-0.25$]{ps}), this size is reached first with teSSM at \unit[$t=(-0.2\pm0.05)$]{ps}, followed by eSSM at \unit[$t=(-0.15\pm0.05)$]{ps} and last with SMI at \unit[$t=(-0.1\pm0.06)$]{ps}. This means that $E_{\textrm{seed}}$ is higher for teSSM, than for eSSM, and lowest for SMI, as expected. As mentioned above, a higher $E_{\textrm{seed}}$ offers several advantages: for example, it reduces the seeding distance $L_\mathrm{seed}$ and allows SM to dominate over other instabilities, such as the beam hose instability~\cite{hosing, moreira}. 

For the bunch and plasma parameters used in the presented measurements, the teSSM configuration increases $E_{\textrm{seed}}$, with respect to eSSM since the available seed bunches are initially longer than optimal (\unit[$\sigma_\mathrm{t,s}\approx2$]{ps} compared to $\sigma_\mathrm{t,s,opt}\sim\unit[1]{ps}$). The teSSM configuration then effectively shortens the seed, removing front-to-back energy transfer in the wakefield and therefore increasing $E_{\textrm{seed}}$ as illustrated in  Fig.~\ref{fig:illustrationteSSM} when comparing $E_{\textrm{seed,a}}$ (Fig.~\ref{fig:illustrationteSSM}a) and $ E_{\textrm{seed,b}}$ (Fig.~\ref{fig:illustrationteSSM}b).

\vspace{0.2cm}

The results in this manuscript experimentally show that truncating seed bunches with a RIF (teSSM) overcomes limitations encountered with established seeding methods~\cite{prllv, IFR}. For RIF seeding, the head of the driver may self-modulate when entering a pre-ionised plasma and seeding with available bunches (Fig.~\ref{Fig2}b) was previously only demonstrated at low plasma density ($n_\mathrm{pe}$)~\cite{prllv}. These limitations are overcome by using teSSM, where seeds are shaped by truncation at the overlap with the RIF. In this way, there are no driver particles ahead of the RIF and the relativistic gamma factor of the seed wakefields is close to that of the drivers, despite evolution. This enables seeding of SM of entire drivers at high $n_\mathrm{pe}$. When using available seed bunches in AWAKE, the timing of micro-bunches is reproducible at the baseline plasma density of \unit[$n_\mathrm{pe}=7\times 10^{14}$]{cm$^{-3}$}, only when using teSSM. Additionally by measuring the time-resolved proton bunch charge density, it can be inferred that teSSM increases the seed wakefield amplitude, which is advantageous since it shortens the length needed until SM saturates and helps SM dominate over other instabilities~\cite{hosing, moreira}. These results are important because they provide a method to seed reproducible SM at any plasma density with available seed bunches. Therefore, the method of teSSM enables controlled high-gradient particle acceleration when requiring SM e.g. for a pre-ionised plasma.

\section{ACKNOWLEDGMENTS}

This work was supported in parts by STFC (AWAKE-UK, Cockcroft Institute core, John Adams Institute core, and UCL consolidated grants), United Kingdom; the National Research Foundation of Korea (Grant No. NRF-2016R1A5A1013277 and No. NRF-2020R1A2C1010835). M. W. acknowledges the support of DESY, Hamburg. Support of the Wigner Datacenter Cloud facility through the Awakelaser project is acknowledged. UW Madison acknowledges support by NSF Award No. PHY-1903316. The AWAKE collaboration acknowledges the SPS team for their excellent proton delivery.

\appendix
\subsection*{Appendix}\label{app1}
For example, for \unit[400]{GeV/c} proton drivers undergoing SM in a plasma of \unit[$n_\mathrm{pe}=7\times 10^{14}$]{cm$^{-3}$} (\unit[$\tau_\mathrm{pe}\approx4$]{ps}) and seeding using \unit[$20$]{MeV} electron bunches (\unit[$\sigma_{\mathrm{z,s}}\approx2$]{ps}, \unit[$n_{b}\approx1.7\times 10^{13}$]{cm$^{-3}$}) with an estimated arrival time jitter of \unit[$\Delta\tau_\mathrm{seed}\sim 1-2$]{ps}:
\begin{enumerate}
    \item \textbf{Dephasing:} Using $\gamma_\mathrm{SM}=427$ and $\gamma_\mathrm{seed}=40$ results in 
    $\Delta\tau / \tau_\mathrm{pe} \sim 1.3~\textrm{ps}/\textrm{m}$.
    Requirement may be violated, depending on $L_\mathrm{seed}$. 
    \item \textbf{Amplitude:} \unit[$E_{\textrm{seed}}\sim 10$]{MV/m}~\cite{prllv} and \unit[$E_{\textrm{noise}}\sim 10$]{kV/m}~\cite{SM}.
    However, the electron bunches exceed the optimal duration (\unit[$\sigma_\mathrm{opt} =\tau_{\mathrm{pe}}/(\sqrt{2}\pi)\sim 1$]{ps}) at this $n_\mathrm{pe}$, reducing $E_{\textrm{seed}}$ through front-to-back energy transfer within the seed.
    \item \textbf{Reproducibility:} Requirement violated since \unit[$\Delta\tau_\mathrm{seed}\sim 1-2$]{ps} and \unit[$\tau_\mathrm{pe}\approx4$]{ps}.

\end{enumerate}
\bibliography{bib}

\end{document}